\documentclass{llncs}

\usepackage[T1]{fontenc}    
\usepackage{hyperref}       
\usepackage{url}            
\usepackage{booktabs}       
\usepackage{amsfonts}       
\usepackage{nicefrac}       
\usepackage{microtype}      
\usepackage{xspace}
\usepackage{booktabs}
\usepackage{subfigure}
\usepackage{placeins}
\usepackage{amssymb}
\usepackage{amsmath}
\usepackage[english]{babel}
\usepackage{color}
\usepackage[dvipsnames]{xcolor}
\definecolor{LightGray}{HTML}{E9E9E9}

\DeclareMathOperator{\argmin}{argmin}

\usepackage{graphicx}
\graphicspath{ {graphics/} }

\usepackage{amsmath}

\newcommand{\eg}{{\textit{e.g.}}}

\newcommand{\etal}{{\textit{et al.}}}
\newcommand{\etc}{{\textit{etc.}}}
\newcommand{\webtotext}{Web2Text\xspace}
\newcommand{\web}{Web\xspace}

\usepackage{soul}
\usepackage{xfrac}

\begin{document}

	\pagestyle{headings}

	\title{\webtotext: Deep Structured Boilerplate Removal}

	\author{Thijs Vogels, Octavian-Eugen Ganea and Carsten Eickhoff}

	\authorrunning{Vogels et al.}

	\institute{Department of Computer Science,\\ETH Zurich,\\Switzerland\\t.vogels@me.com, octavian.ganea@inf.ethz.ch, c.eickhoff@acm.org}

	\maketitle

	\begin{abstract}
		Web pages are a valuable source of information for many natural language processing and information retrieval tasks. Extracting the main content from those documents is essential for the performance of derived applications. To address this issue, we introduce a novel model that performs sequence labeling to collectively classify all text blocks in an HTML page as either boilerplate or main content. Our method uses a hidden Markov model on top of potentials derived from DOM tree features using convolutional neural networks. The proposed method sets a new state-of-the-art performance for boilerplate removal on the CleanEval benchmark. As a component of information retrieval pipelines, it improves retrieval performance on the ClueWeb12 collection.
	\end{abstract}

	\section{Introduction}\label{sec:intro}
	Modern methods in natural language processing and information retrieval are heavily dependent on large collections of text. The World Wide Web is an inexhaustible source of content for such applications. However, a common problem is that \web pages include not only main content, but also ads, hyperlink lists, navigation, previews of other articles, banners, \textit{etc}. This boilerplate/template content has often been shown to have negative effects on the performance of derived applications~\cite{kohlschutter2009densitometric,yi2003eliminating}.

	The task of separating main text in a \web page from the remaining content is known in the literature as ``boilerplate removal'', ``\web page segmentation'' or ``content extraction''. Established popular methods for this problem use rule-based or machine learning algorithms. The most successful approaches first perform a splitting of an input \web page into text blocks, followed by a binary labeling of each block as either main content or boilerplate.

	In this paper, we propose a hidden Markov model on top of neural potentials for the task of boilerplate removal. We leverage the representational power of convolutional neural networks (CNNs) to learn unary and pairwise potentials over blocks in a page-based on complex non-linear combinations of DOM-based traditional features. At prediction time, we find the most likely block labeling by maximizing the joint probability of a label sequence using the Viterbi algorithm~\cite{viterbi2010error}. The effectiveness of our method is demonstrated on standard benchmarking datasets.

	The remainder of this document is structured as follows. Section~\ref{sec:related} gives an overview of related work. Section~\ref{sec:model} formally defines the main-content extraction problem, introduces the block segmentation procedure and details our model. Section~\ref{sec:experiments} empirically demonstrates the merit of our method on several benchmark datasets for content extraction and document retrieval.


	\section{Related Work}\label{sec:related}
	Early approaches to HTML boilerplate removal use a range of heuristics and rule-based methods. Finn \etal~\cite{finn2001fact} design an effective system called \emph{Body Text Extractor} (BTE). It relies on the observation that the main content contains longer paragraphs of uninterrupted text, where HTML tags occur less frequently compared to the rest of the \web page.
	Looking at the cumulative distribution of tags as a function of the position in the document, Finn \etal\ identify a flat region in the middle of this distribution graph to be the main content of the page. While simple, their algorithm has two drawbacks: (1) it only makes use of the location of HTML tags and not of their structure, thus losing potentially valuable information, and (2) it can only identify one continuous stretch of main content which is unrealistic for a considerable percentage of modern \web pages.

	To address these issues, several other algorithms have been designed to operate on DOM trees, thus leveraging the semantics of the HTML structure~\cite{gupta2003dom,lin2002discovering,debnath2005automatic}. The problem with these early methods is that they make intensive use of the fact that pages used to be partitioned into sections by \texttt{<table>} tags, which is no longer a valid assumption.

	In the next line of work, the DOM structure is used to jointly process multiple pages from the same domain, relying on their structural similarities. This approach was pioneered by Yi \etal~\cite{yi2003eliminating} and was improved by various others~\cite{vieira2006fast}. These methods are very suitable for detecting template content that is present in all pages of a website, but have poor performance on websites that consist of a single \web page only. In this paper we focus on single-page content extraction without exploiting the context of other pages from the same site.

	Gottron \etal~\cite{gottron2008content} propose \emph{Document Slope Curves} and \emph{Content Code Blurring} methods that are able to identify multiple disconnected content regions. The latter method parses the HTML source code as a vector of 1's, representing pieces of text, and 0's, representing tags. This vector is then smoothed iteratively, such that eventually it finds active regions where text dominates (content) and inactive regions where tags dominate (boilerplate). This idea of smoothing was extended to also deal with the DOM structure~\cite{chakrabarti2008graph,sun2011dom}. Chakrabarti \etal~\cite{chakrabarti2007page} assign a likelihood of being content to each leaf of the DOM tree while using isotonic smoothing to combine the likelihoods of neighbors with the same parents. In a similar direction, Sun \etal~\cite{sun2011dom} use both the tag/text ratio and DOM tree information to propagate \emph{DensitySums} through the tree.

	Machine learning methods offer a convenient way to combine various indicators of ``contentness'', automatically weighting hand-crafted features according to their relative importance. The FIASCO system by Bauer \etal~\cite{bauer2007fiasco} uses Support Vector Machines (SVM) to classify an HTML page as a sequence of blocks that are generated through a DOM-based segmentation of the page and are represented by linguistic, structural and visual features. Similar works of Kohlsch\"{u}tter \etal~\cite{kohlschutter2010boilerplate} also employ SVMs to independently classify blocks.
	Spousta et al.~\cite{spousta2008victor} extend this approach by reformulating the classification problem as a case of sequence labeling where all blocks are jointly tagged. They use conditional random fields to take advantage of correlations between the labels of neighboring content blocks. This method was the most successful in the CleanEval competition~\cite{baroni2008cleaneval}.

	In this paper, we propose an effective set of block features that capture information from adjacent neighbors in the DOM tree. Additionally, we employ a deep learning framework to automatically learn non-linear features combinations, giving the model an advantage over traditional linear approaches. Finally, we jointly optimize the labels for the whole \web page according to local potentials predicted by the neural networks.

	\section{\webtotext}\label{sec:model}
\begin{figure}[t]
		\centering
		\includegraphics[width=\columnwidth]{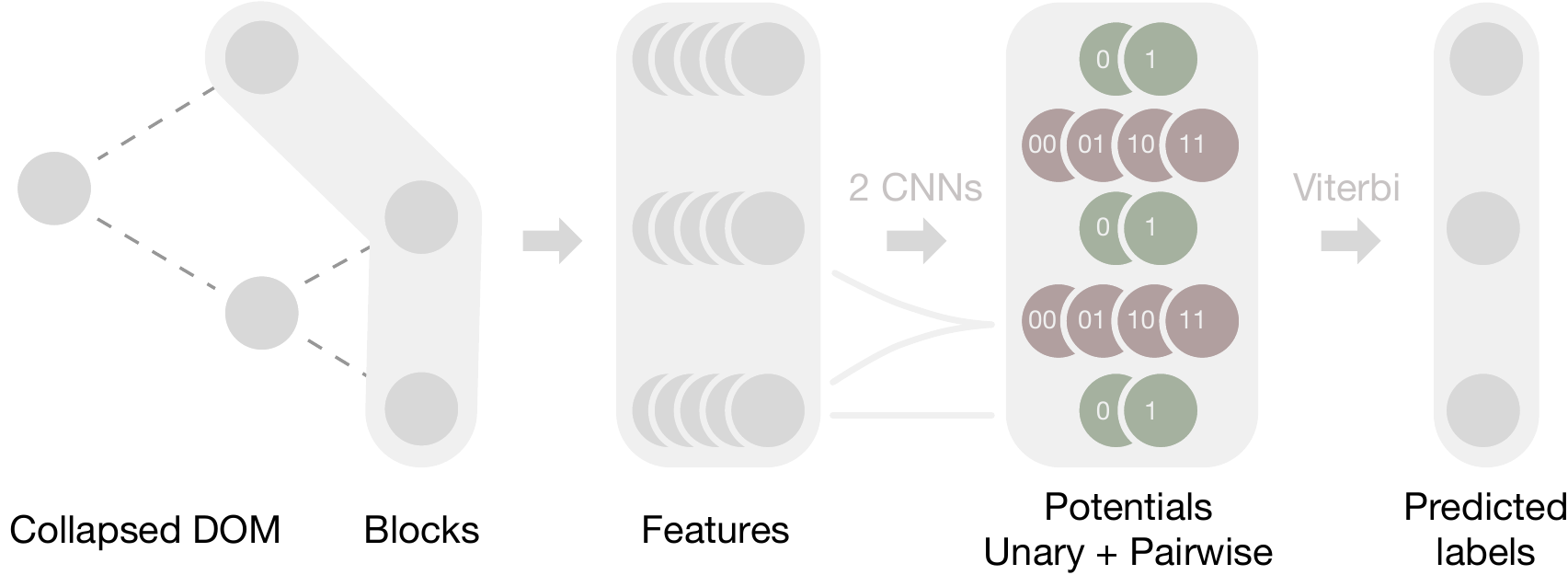}
		\caption{The \webtotext pipeline. The leaves of the Collapsed DOM tree of a \web page form an ordered sequence of blocks to be labeled. For each block, we extract a number of DOM tree-based features. Two separate convolutional networks operating on this sequence of features yield two respective sets of potentials: unary potentials for each block and pairwise potentials for each pair of neighboring blocks. These define a hidden Markov model. Using the Viterbi algorithm, we find an optimal labeling that maximizes the total sequence probability as predicted by the neural networks.}
		\label{fig:overview}
	\end{figure}

	Boilerplate removal is the problem of labeling sections of the text of a \web page as \emph{main content} or \emph{boilerplate} (anything else)~\cite{baroni2008cleaneval}. In the following, we discuss the various steps of our method. The complete pipeline is also illustrated in Figure~\ref{fig:overview}.

	\subsection{Preprocessing}
	We expect raw \web page input to be written in (X)HTML markup. Each document is parsed as a Document Object Model tree (DOM tree) using Jsoup~\cite{jsoup}. We preprocess this DOM tree by i) removing empty nodes or nodes containing only whitespace, ii) removing nodes that do not have any content we can extract: \eg\ \texttt{<br>}, \texttt{<checkbox>}, \texttt{<head>}, \texttt{<hr>}, \texttt{<iframe>}, \texttt{<img>}, \texttt{<input>}.
	
	\begin{figure}[t]
		\centering
		\includegraphics[width=\columnwidth]{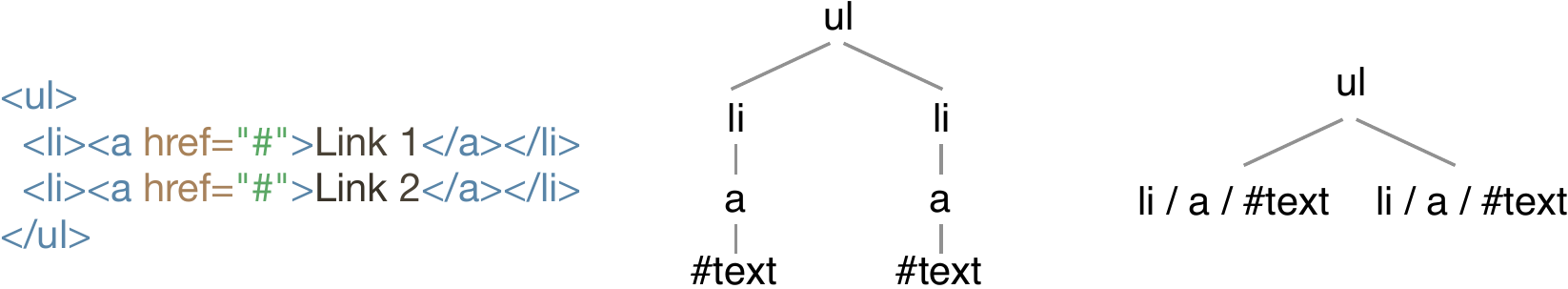}
		\caption{Collapsed DOM procedure example. \emph{Left}: HTML source code, \emph{middle}: the corresponding DOM tree, \emph{right}: the corresponding Collapsed DOM.}
		\label{fig:cdom}
	\end{figure}

	We make use of the parent and grandparent DOM tree relations. In a raw DOM tree, however, these relationships are not always meaningful. Figure~\ref{fig:cdom} shows a typical fragment of a DOM tree where two neighboring nodes share the same semantic parent (\texttt{<ul>}) but not the same DOM parent. To improve the expressiveness of tree based features (such as ``the number of children of a node's parent''), we recursively merge single child parent nodes with their respective child. We call the resulting tree-structure the \emph{Collapsed~DOM} (CDOM). 


	\subsection{Block Segmentation} \label{sec:segmentation}

	Our content extraction algorithm is based on sequence labeling. A \web page is treated as a sequence of blocks that are labeled \emph{main content} or \emph{boilerplate}. There are multiple ways to split a \web page into blocks, the most popular currently used being i) Lines in the HTML file, ii) DOM leaves, iii) Block-level DOM leaves. We opt for using the most flexible \textit{DOM leaves} strategy, described as follows. Sections on a page that require different labels are usually separated by at least one HTML tag. Therefore, it is safe to consider DOM leaves (\texttt{\#text} nodes) as the blocks of our sequence. A potential disadvantage of this approach is that a hyperlink in a text paragraph can receive a different label than its neighboring text. Under this scheme, an empirical evaluation of \webtotext shows no cases where parts of a textual paragraph are wrongly labeled as \emph{boilerplate}, while the rest are marked as \emph{main content}.

%
%
%


	\subsection{Feature Extraction}\label{feature}

	Features are properties of a node that may be indicative of it being content or boilerplate. Such features can be based on the node's text, CDOM structure or a combination thereof. We distinguish between block features and edge features.

	\emph{Block features} capture information on each block of text on a page. They are statistics collected based on the block's CDOM node, parent node, grandparent node and the root of the CDOM tree. In total, we collect 128 features for each text block, \eg\ ``the node is a \texttt{<p>} element'', ``average word length'', ``relative position in the source code'', ``the parent node's text contains an email address'', ``ratio of stopwords in the whole page'', \etc\ We clip and standardize all non-binary features to be approximately Gaussian with zero mean and unit variance across the training set. For a full overview of all 128 features, please refer to Appendix~\ref{sec:block-features}.

	\emph{Edge features} capture information on each pair of neighboring text blocks. We collect 25 features for each such pair. Define the \emph{tree distance} of two nodes as the sum of the number of hops from both nodes to their first common ancesor. The first edge features we use are binary features corresponding to a tree distance of 2, 3, 4 and $>4$. Another feature signifies if there is a \emph{line break} between the nodes in an unstyled HTML page. Finally, we collect features b70--b89 from Appendix~\ref{sec:block-features} for the \emph{common ancestor} CDOM node of the two text blocks.


	\subsection{CNN Unary and Pairwise Potentials}

	We assign unary potentials to each text block to be labeled and pairwise potentials to each pair of neighboring text blocks. In our case, potentials are probabilities as explained below. The unary potentials $p_i(l_i = 1)$, $p_i(l_i = 0)$ are the probabilities that the label $l_i$ of a text block $i$ is content or boilerplate, respectively. The two potentials sum to one. The pairwise potentials $p_{i,i+1}(l_i = 1, l_{i+1}=1)$, $p_{i,i+1}(l_i = 1, l_{i+1}=0)$, $p_{i,i+1}(l_i = 0, l_{i+1}=1)$ and $p_{i,i+1}(l_i = 0, l_{i+1}=0)$ are the transition probabilities of the labels of a pair of neighboring text blocks. These pairwise potentials also sum to one for each text block pair.

	The two sets of potentials are modeled using CNNs with 5 layers, ReLU non-linearity between layers, filter sizes of $(50, 50, 50, 10, 2)$ for the unary network and of $(50, 50, 50, 10, 4)$ for the pairwise network. All filters have a stride of 1 and kernel sizes $(1, 1, 3, 3, 3)$ respectively. The unary CNN receives a sequence of block features corresponding to the sequence of text blocks to be labeled and outputs unary potentials for each block. The pairwise CNN receives a sequence of edge features corresponding to the sequence of edges to be labeled and outputs the pairwise potentials for each block. We use zero padding to make sure that each layer produces a sequence of the same size as its input sequence. The outputs for the unary network are sequences of 2 values per block that are normalized using \text{softmax}. The outputs for the pairwise network are sequences of 4 values per block-pair that are normalized in the same way. Thus, the output for the block $i$ depends indirectly on a range of blocks around it. We employ dropout regularization with rate 0.2 and $L_2$ weight decay with rate $10^{-4}$.

	For the unary potentials, we minimize the cross-entropy
\begin{align}
		\vec{\theta}_{\text{unary}}^* = \argmin_{\vec{\theta}_{\text{unary}}} -\sum_{i=0}^n \log p_i(l_i=l_i^* \;|\; \vec{\theta}_{\text{unary}}),
\end{align}
	where $l_i^*$ is the true label of block $i$, $\vec{\theta}_{\text{unary}}$ are the parameters of the unary network and $n$ is the index of the last text block in the sequence. 
	
	For the pairwise network, we minimize the cross-entropy
\begin{align}
	\vec{\theta}_{\text{pairwise}}^* = \argmin_{\vec{\theta}_{\text{pairwise}}} -\sum_{i=0}^{n-1} \log p_{i(i+1)}(l_i=l_i^*,\;l_{i+1}=l_{i+1}^* \;|\; \vec{\theta}_{\text{pairwise}}),
\end{align}
	where $\vec{\theta}_\text{pairwise}$ are the parameters of the pairwise network.


	\subsection{Inference}
	The joint prediction of the most likely sequence of labels given an input \web page works as follows. We denote the sequence of text blocks on the page as $(b_0,b_1,\ldots,b_n)$ and write the probability of a corresponding labeling $(\ell_0,\ell_1,\ldots,\ell_n) \in \{\text{0},\text{1}\}^n$ being the correct one as
\begin{align}
	p(\ell_0,\ldots,\ell_n) = \left(\prod_{i=0}^n p_i(\ell_i) \right) \left(  \prod_{i=0}^{n-1} p_{i(i+1)}(\ell_{i}, \ell_{i+1})\right)^\lambda,
\end{align}
	where $\lambda$ is an interpolation factor between the unary and pairwise terms. We use $\lambda=0.1$ in our experiments.
	This expression describes a hidden Markov model and it is maximized using the Viterbi algorithm~\cite{viterbi2010error} to find the optimal labeling given the predicted CNN potentials.







	\section{Experiments}\label{sec:experiments}

	Our experiments are grouped in two stages. We begin by assessing \webtotext's performance at boilerplate removal on a high-quality manually annotated corpus of \web pages. In a second step, we turn towards a much larger collection and investigate how improved content extraction results in superior information retrieval quality. Both experiments highlight the benefits of \webtotext over state-of-the-art alternatives.

	\subsection{Training Data}\label{sec:train-data}
	CleanEval 2007~\cite{baroni2008cleaneval} is the largest publicly available dataset for this task. It contains 188 text blocks per Web page on average. It consists of an original split of development (60 pages) and test (676 pages) sets. We divide the development set into a training set (55 pages) and a test set (5 pages). Since our model has more than 10,000 parameters, it is likely that the original training set is too small for our method. Thus, we did a second split of the CleanEval as follows: training (531 pages), validation (58 pages) and test (148 pages). 

	\subsubsection{Automatic Block Labeling.}\label{sssec:dp-labeling}

	To our knowledge, the existing corpora (including CleanEval) for boilerplate detection pose an additional difficulty. These datasets consist only of pairs of \web pages and corresponding cleaned text (manually extracted). As a consequence, the alignment between the source text and cleaned text, as well as block labeling, have to be recovered. Some methods (\eg~\cite{spousta2008victor}) rely on expensive manual block annotations. One of our contributions is the following automatic recovery procedure of the aligned (block, label) pairs from the original (\web page, clean text) pairs. This allows us to leverage more training data compared to previous methods.
	
	We first linearly scan the cleaned text of a \web page using windows of 10 consecutive characters. Each such snippet is checked for uniqueness in the original \web page (after spaces trimming). If such a unique match is found, then it can be used to divide both the cleaned text and the original \web page in two parts on which the same matching method can be applied recursively in a \textit{divide-et-impera} fashion. After all unique snippets are processed, we use dynamic programming to align the remaining splitted parts of the clean text with the corresponding splitted parts of the original \web page blocks. In the end, in the rare case that the content of a block is only partially matched with the cleaned text, we mark it as \textit{content} iff at least \sfrac{2}{3} of its text is aligned.

	\subsection{Training Details}
  The unary and pairwise potential-predicting networks are trained separately with the Adam optimizer \cite{kingma2014adam} and a learning rate of $10^{-3}$ for 5000 iterations. Each iteration processes a mini-batch of $128$ 9-text-block long \web page excerpts. We perform early stopping, observing no improvements after this number of steps. We then pick the model corresponding to the lowest error on the validation set.


  \subsection{Baselines}
  We compare \webtotext to a range of methods described in the literature or deployed in popular libraries. BTE~\cite{finn2001fact} and Unfluff~\cite{geitgey2014unfluff} are heuristic methods. \cite{kohlschutter2010boilerplate,kohlschutter2010boilerpipe} is a popular machine learning system that offers various content extraction settings\footnote{We were not able to find code for re-training this system.} which we used in our experiments (see Table~\ref{table:results}). CRF~\cite{spousta2008victor} achieves one of the best results on CleanEval. This machine learning model trains a Conditional Random Field on top of block features in order to perform block classification. However, as explained in Section~\ref{sssec:dp-labeling}, CRF relies on a different \web page block splitting and on expensive manual block annotations. As a consequence, we were not able to re-train it and thus only used their out-of-the-box model pre-trained on the original CleanEval split. For a fair comparison, we also train on the original CleanEval split, but note below that our neural network has many more parameters and will suffer from using so few training instances.
  
  \subsubsection{Model Sizes.} The CRF model~\cite{spousta2008victor} contains 9,705 parameters. In comparison, our unary CNN network contains 17,960 parameters, while the pairwise CNN contains 12,870 parameters, the total number of parameters for the joint structured model being 30,830. This explains why the original train set is too small for our model.

	\setlength{\tabcolsep}{2.5pt}
	\begin{table}[t]
		\centering
		\caption{Boilerplate removal results on the CleanEval dataset. We use two different splits of this dataset, the original split (55p, 5p, 676p) and our split (531p, 58p, 148p). It is confirmed that our method benefits from bigger training sets. \label{table:results}}
		\centerline{
			\begin{tabular}{ l  c c c c  c  c c c c}
          & \multicolumn{4}{c}{\begin{tabular}{@{}c@{}}Original test (676 pages) \end{tabular}} & & \multicolumn{4}{c}{\begin{tabular}{@{}c@{}}Our test (148 pages) \end{tabular}} \\\cmidrule{2-5}\cmidrule{7-10}
				Method & Acc. & Precision & Recall & $F_1$ & & Acc. & Precision & Recall & $F_1$\\
				\hline
				\begin{tabular}{@{}l@{}}CRF~\cite{spousta2008victor} \\ original train 55p + 5p\end{tabular} & 0.82 & 0.87 & 0.81 & 0.84 & & 0.82 & 0.88 & 0.81 & 0.84\\
				BTE~\cite{finn2001fact} & 0.79 & 0.79 & \textbf{0.89} & 0.83 & & 0.75 & 0.76 & 0.84 & 0.80\\
				default-ext~\cite{kohlschutter2010boilerpipe} &  0.80 & 0.89 & 0.75 & 0.81 & & 0.79 & 0.89 & 0.74 & 0.81\\
				article-ext~\cite{kohlschutter2010boilerpipe} &  0.72 & 0.91 & 0.59 & 0.71 & & 0.67 & 0.89 & 0.50 & 0.64\\
				largest-ext~\cite{kohlschutter2010boilerpipe} & 0.60 & \textbf{0.93} & 0.36 & 0.52 & & 0.59 &\textbf{0.93} & 0.33 & 0.48\\
				Unfluff~\cite{geitgey2014unfluff} & 0.71 & 0.90 & 0.57 & 0.70 & & 0.68 & 0.90 & 0.51 & 0.65\\
				\begin{tabular}{@{}l@{}} \textbf{\webtotext} \\ original train 55p, val 5p\end{tabular} & \textbf{0.84} & 0.88 & 0.85 & \textbf{0.86}\\
				\begin{tabular}{@{}l@{}} \textbf{\webtotext} \\ our train 531p, val 58p\end{tabular} & & & & & & \textbf{0.86} & 0.87 & \textbf{0.90} & \textbf{0.88}\\
				\hline
		\end{tabular}}
	\end{table}
	
	\subsection{Content Extraction Results}
	Table~\ref{table:results} shows the results of this experiment. All the metrics are block based, where all blocks are weighted equally. We note that \webtotext obtains state-of-the-art accuracy, recall and F1 scores compared to popular baselines including previous CleanEval winners. Note that these numbers are obtained by evaluating each method using the same block segmentation procedure, namely the DOM leaves strategy described in Section~\ref{sec:segmentation}. We additionally note that, compared to using \webtotext  only with the unary CNN, the gains of the hidden Markov model are marginal in this experiment.

	\subsubsection{Running times.} \webtotext takes 54ms per \web page on average; 35ms for DOM parsing and feature extraction, and 19ms for the neural network forward pass and Viterbi algorithm. These measurements were done on a Macbook with a 2.8 GHz Intel Core i5 processor.

	\begin{table*}[!h]
		\centering
		\caption{The effect of boilerplate removal on \textit{ad hoc} retrieval performance. An asterisk (*) indicates a significance performance difference between raw and cleaned HTML concent. A dagger ($\dag$) indicates that a model significantly outperforms all other text extraction methods.\label{tab:retrieval}}
			\begin{tabular}{ l c l l l l l l }
				Collection & Ret.\ Model & Method & P@10 & R@10 & $F_1$@10 & MAP & nDCG\\
				\hline
				CW12-A & QL & raw content & 0.316 & 0.056 & 0.095 & 0.137 & 0.459\\
				CW12-A & QL & CRF~\cite{spousta2008victor} & 0.342* & 0.068* & 0.113* & 0.147* & 0.543*\\
				CW12-A & QL & BTE~\cite{finn2001fact} & 0.301 & 0.048 & 0.083 & 0.128 & 0.435\\
				CW12-A & QL & default-ext~\cite{kohlschutter2010boilerpipe} & 0.318 & 0.055 & 0.094 & 0.138 & 0.462\\
				CW12-A & QL & article-ext~\cite{kohlschutter2010boilerpipe} & 0.298 & 0.049 & 0.084 & 0.126 & 0.433\\
				CW12-A & QL & largest-ext~\cite{kohlschutter2010boilerpipe} & 0.279 & 0.044 & 0.076 & 0.112 & 0.417\\
				CW12-A & QL & Unfluff~\cite{geitgey2014unfluff} & 0.304 & 0.051 & 0.087 & 0.128 & 0.428\\
				CW12-A & QL & \webtotext & \textbf{0.361*$^\dag$} & \textbf{0.079*$^\dag$} & \textbf{0.130*$^\dag$} & \textbf{0.154*$^\dag$} & \textbf{0.578*$^\dag$}\\
				\hline
				CW12-A & RM & raw content & 0.278 & 0.048 & 0.082 & 0.121 & 0.439\\
				CW12-A & RM & CRF~\cite{spousta2008victor} & 0.301* & 0.057* & 0.096* & 0.138* & 0.487*\\
				CW12-A & RM & BTE~\cite{finn2001fact} & 0.262 & 0.041 & 0.071 & 0.110 & 0.409\\
				CW12-A & RM & default-ext~\cite{kohlschutter2010boilerpipe} & 0.277 & 0.048 & 0.082 & 0.123 & 0.442\\
				CW12-A & RM & article-ext~\cite{kohlschutter2010boilerpipe} & 0.260 & 0.039 & 0.068 & 0.109 & 0.411\\
				CW12-A & RM & largest-ext~\cite{kohlschutter2010boilerpipe} & 0.248 & 0.032 & 0.057 & 0.097 & 0.401\\
				CW12-A & RM & Unfluff~\cite{geitgey2014unfluff} & 0.264 & 0.041 & 0.071 & 0.111 & 0.407\\
				CW12-A & RM & \webtotext & \textbf{0.325*$^\dag$} & \textbf{0.069*$^\dag$} & \textbf{0.114*$^\dag$} & \textbf{0.145*$^\dag$} & \textbf{0.525*$^\dag$}\\
				\hline
				CW12-B & QL & raw content & 0.210 & 0.025 & 0.045 & 0.037 & 0.134\\
				CW12-B & QL & CRF~\cite{spousta2008victor} & 0.241* & 0.031* & 0.055* & 0.048* & 0.165*\\
				CW12-B & QL & BTE~\cite{finn2001fact} & 0.193 & 0.019 & 0.035 & 0.030 & 0.121\\
				CW12-B & QL & default-ext~\cite{kohlschutter2010boilerpipe} & 0.212 & 0.026 & 0.046 & 0.038 & 0.132\\
				CW12-B & QL & article-ext~\cite{kohlschutter2010boilerpipe} & 0.199 & 0.017 & 0.031 & 0.031 & 0.120\\
				CW12-B & QL & largest-ext~\cite{kohlschutter2010boilerpipe} & 0.178 & 0.015 & 0.028 & 0.024 & 0.107\\
				CW12-B & QL & Unfluff~\cite{geitgey2014unfluff} & 0.195 & 0.020 & 0.036 & 0.029 & 0.121\\
				CW12-B & QL & \webtotext & \textbf{0.266*$^\dag$} & \textbf{0.038*$^\dag$} & \textbf{0.067*$^\dag$} & \textbf{0.055*$^\dag$} & \textbf{0.181*$^\dag$}\\
				\hline
				CW12-B & RM & raw content & 0.172 & 0.021 & 0.037 & 0.030 & 0.122\\
				CW12-B & RM & CRF~\cite{spousta2008victor} & 0.198* & 0.028* & 0.049* & 0.041* & 0.143*\\
				CW12-B & RM & BTE~\cite{finn2001fact} & 0.158 & 0.015 & 0.027 & 0.022 & 0.111\\
				CW12-B & RM & default-ext~\cite{kohlschutter2010boilerpipe} & 0.170 & 0.020 & 0.036 & 0.029 & 0.124\\
				CW12-B & RM & article-ext~\cite{kohlschutter2010boilerpipe} & 0.156 & 0.015 & 0.027 & 0.019 & 0.109\\
				CW12-B & RM & largest-ext~\cite{kohlschutter2010boilerpipe} & 0.145 & 0.013 & 0.024 & 0.015 & 0.095\\
				CW12-B & RM & Unfluff~\cite{geitgey2014unfluff} & 0.159 & 0.016 & 0.029 & 0.021 & 0.112\\
				CW12-B & RM & \webtotext & \textbf{0.213*$^\dag$} & \textbf{0.032*$^\dag$} & \textbf{0.056*$^\dag$} & \textbf{0.046*$^\dag$} & \textbf{0.165*$^\dag$}\\
				\hline
		\end{tabular}
	\end{table*}

	\subsection{Impact on Retrieval Performance}

	Besides the previously presented intrinsic evaluation of text extraction accuracy, we are interested in the performance gains that other derived tasks experience when operating on the output of boilerplate removal systems of varying quality. To this end, our extrinsic evaluation studies the task of \textit{ad hoc} document retrieval. Search engines that index high-quality output of text extraction systems should be better able to answer a given user-formulated query than systems indexing raw HTML or na\"{i}vely cleaned content. Our experiments are based on the well-known ClueWeb12 collection of \web pages.\footnote{http://lemurproject.org/clueweb12/} It is organized in two well-defined document sets, the full CW12-A corpus of 733M organic \web documents (27.3 TB of uncompressed text) as well as the smaller, randomly sampled subset CW12-B of 52M documents (1.95 TB of uncompressed text). The collection is indexed using the Indri search engine and retrieval runs are conducted using two state-of-the-art probabilistic retrieval models, the query likelihood model~\cite{jin2002language} (QL) as well as a relevance-based language model~\cite{lavrenko2001relevance} (RM). Our 50 test queries alongside their relevance judgments originate from the 2013 edition of the TREC Web Track~\cite{collins2013overview}.

	Table~\ref{tab:retrieval} highlights the quality of each combination of retrieval model and collection when indexing either raw or cleaned \web content. Within each combination, statistical significance of performance differences between raw and cleaned HTML content is denoted by an asterisk. Models that significantly outperform all other text extraction methods are indicated by $\dag$. We can note that, in general, retrieval systems indexing CW12-A deliver stronger results than those operating only on the CW12-B subset. Due to the random sampling process, many potentially relevant documents are missing from this smaller collection. Similarly, across all comparable settings, the query likelihood model (QL) performs significantly better than the relevance model (RM). As hypothesized earlier, text extraction can influence the quality of subsequent document retrieval. We note that low-recall methods (BTE, article-ext, largest-ext, Unfluff) cause losses in retrieval performance, as relevant pieces of content are incorrectly removed as boilerplate. At the same time, the most accurate models (CRF, \webtotext) were able to introduce improvements across all metrics. \webtotext, in particular, outperformed all baselines at significance level $0.05$. We note that, for this experiment, \webtotext was trained on our CleanEval split as explained in Section~\ref{sec:train-data}.


	\section{Conclusion}\label{sec:conclusion}
	This paper presents \webtotext\footnote{Our source code is publicly available: https://github.com/dalab/web2text}, a novel algorithm for main content extraction from \web pages. The method combines the virtues of popular sequence labeling approaches such as CRFs~\cite{gibson2007adaptive} with deep learning methods that leverage the DOM structure as a source of information. Our experimental evaluation on CleanEval benchmarking data shows significant performance gains over all state-of-the-art methods. In a second set of experiments, we demonstrate how highly accurate boilerplate removal can significantly increase the performance of derived tasks such as \emph{ad hoc} retrieval.

  \section*{Acknowledgments}
  This research is funded by the Swiss National Science Foundation (SNSF) under grant agreement numbers 167176 and 174025.

	\bibliographystyle{plain}
	\bibliography{ref.bib}

\begin{thebibliography}{10}

\bibitem{baroni2008cleaneval}
Marco Baroni, Francis Chantree, Adam Kilgarriff, and Serge Sharoff.
\newblock {CleanEval}: a competition for cleaning web pages.
\newblock In {\em LREC}, 2008.

\bibitem{bauer2007fiasco}
Daniel Bauer, Judith Degen, Xiaoye Deng, Priska Herger, Jan Gasthaus, Eugenie
  Giesbrecht, Lina Jansen, Christin Kalina, Thorben Kr\"ager, Robert M\"artin,
  Martin Schmidt, Simon Scholler, Johannes Steger, Egon Stemle, and Stefan
  Evert.
\newblock {FIASCO}: Filtering the internet by automatic subtree classification,
  osnabruck.
\newblock In {\em Building and Exploring Web Corpora: Proceedings of the 3rd
  Web as Corpus Workshop, incorporating {CleanEval}}, volume~4, pages 111--121,
  2007.

\bibitem{chakrabarti2007page}
Deepayan Chakrabarti, Ravi Kumar, and Kunal Punera.
\newblock Page-level template detection via isotonic smoothing.
\newblock In {\em Proceedings of the 16th international conference on World
  Wide Web}, pages 61--70. ACM, 2007.

\bibitem{chakrabarti2008graph}
Deepayan Chakrabarti, Ravi Kumar, and Kunal Punera.
\newblock A graph-theoretic approach to webpage segmentation.
\newblock In {\em Proceedings of the 17th international conference on World
  Wide Web}, pages 377--386. ACM, 2008.

\bibitem{collins2013overview}
Kevyn Collins-Thompson, Paul Bennett, Fernando Diaz, Charlie Clarke, and Ellen
  Voorhees.
\newblock Overview of the {TREC} 2013 web track.
\newblock In {\em Proceedings of the 22nd Text Retrieval Conference (TREC'13)},
  2013.

\bibitem{debnath2005automatic}
Sandip Debnath, Prasenjit Mitra, Nirmal Pal, and C~Lee Giles.
\newblock Automatic identification of informative sections of web pages.
\newblock {\em IEEE transactions on knowledge and data engineering},
  17(9):1233--1246, 2005.

\bibitem{finn2001fact}
Aidan Finn, Nicholas Kushmerick, and Barry Smyth.
\newblock Fact or fiction: Content classification for digital libraries.
\newblock {\em Unrefereed}, 2001.

\bibitem{geitgey2014unfluff}
Adam Geitgey.
\newblock Unfluff -- an automatic web page content extractor for node.js!,
  2014.

\bibitem{gibson2007adaptive}
John Gibson, Ben Wellner, and Susan Lubar.
\newblock Adaptive web-page content identification.
\newblock In {\em Proceedings of the 9th annual ACM international workshop on
  Web information and data management}, pages 105--112. ACM, 2007.

\bibitem{gottron2008content}
Thomas Gottron.
\newblock Content code blurring: A new approach to content extraction.
\newblock In {\em Database and Expert Systems Application, 2008. DEXA'08. 19th
  International Workshop on}, pages 29--33. IEEE, 2008.

\bibitem{gupta2003dom}
Suhit Gupta, Gail Kaiser, David Neistadt, and Peter Grimm.
\newblock {DOM}-based content extraction of {HTML} documents.
\newblock In {\em Proceedings of the 12th international conference on World
  Wide Web}, pages 207--214. ACM, 2003.

\bibitem{jsoup}
Jonathan Hedley.
\newblock Jsoup {HTML} parser, 2009.

\bibitem{jin2002language}
Rong Jin, Alex~G Hauptmann, and ChengXiang Zhai.
\newblock Language model for information retrieval.
\newblock In {\em Proceedings of the 25th annual international ACM SIGIR
  conference on Research and development in information retrieval}, pages
  42--48. ACM, 2002.

\bibitem{kingma2014adam}
Diederik Kingma and Jimmy Ba.
\newblock Adam: A method for stochastic optimization.
\newblock {\em arXiv preprint arXiv:1412.6980}, 2014.

\bibitem{kohlschutter2009densitometric}
Christian Kohlsch{\"u}tter.
\newblock A densitometric analysis of web template content.
\newblock In {\em Proceedings of the 18th international conference on World
  wide web}, pages 1165--1166. ACM, 2009.

\bibitem{kohlschutter2010boilerpipe}
Christian Kohlsch\"{u}tter et~al.
\newblock Boilerpipe -- boilerplate removal and fulltext extraction from {HTML}
  pages.
\newblock {\em Google Code}, 2010.

\bibitem{kohlschutter2010boilerplate}
Christian Kohlsch{\"u}tter, Peter Fankhauser, and Wolfgang Nejdl.
\newblock Boilerplate detection using shallow text features.
\newblock In {\em Proceedings of the third ACM international conference on Web
  search and data mining}, pages 441--450. ACM, 2010.

\bibitem{lavrenko2001relevance}
Victor Lavrenko and W~Bruce Croft.
\newblock Relevance based language models.
\newblock In {\em Proceedings of the 24th annual international ACM SIGIR
  conference on Research and development in information retrieval}, pages
  120--127. ACM, 2001.

\bibitem{lin2002discovering}
Shian-Hua Lin and Jan-Ming Ho.
\newblock Discovering informative content blocks from web documents.
\newblock In {\em Proceedings of the eighth ACM SIGKDD international conference
  on Knowledge discovery and data mining}, pages 588--593. ACM, 2002.

\bibitem{spousta2008victor}
Miroslav Spousta, Michal Marek, and Pavel Pecina.
\newblock Victor: the web-page cleaning tool.
\newblock In {\em 4th Web as Corpus Workshop (WAC4)-Can we beat Google}, pages
  12--17, 2008.

\bibitem{sun2011dom}
Fei Sun, Dandan Song, and Lejian Liao.
\newblock Dom based content extraction via text density.
\newblock In {\em Proceedings of the 34th international ACM SIGIR conference on
  Research and development in Information Retrieval}, pages 245--254. ACM,
  2011.

\bibitem{vieira2006fast}
Karane Vieira, Altigran~S Da~Silva, Nick Pinto, Edleno~S De~Moura, Joao
  Cavalcanti, and Juliana Freire.
\newblock A fast and robust method for web page template detection and removal.
\newblock In {\em Proceedings of the 15th ACM international conference on
  Information and knowledge management}, pages 258--267. ACM, 2006.

\bibitem{viterbi2010error}
Andrew~J Viterbi.
\newblock Error bounds for convolutional codes and an asymptotically optimum
  decoding algorithm.
\newblock In {\em The Foundations Of The Digital Wireless World: Selected Works
  of AJ Viterbi}, pages 41--50. World Scientific, 2010.

\bibitem{yi2003eliminating}
Lan Yi, Bing Liu, and Xiaoli Li.
\newblock Eliminating noisy information in web pages for data mining.
\newblock In {\em Proceedings of the ninth ACM SIGKDD international conference
  on Knowledge discovery and data mining}, pages 296--305. ACM, 2003.

\end{thebibliography}
	\FloatBarrier

  \newpage	
	\appendix
	\section{List of block features} \label{sec:block-features}
	\begin{table}[!h]
		\caption{List of all block features used. 1/0 indicated a binary feature: 1 if true, 0 if false. Non-binary features are normalized to have zero mean and unit variance.}
	
		\centering 
		\makebox[\textwidth]{
		\def\arraystretch{1.06}
		\begin{tabular}{lll}
		ID & Name & Description \\
		\hline
		b1         & has duplicate              & 1/0: is there another node with the same text?                                                                                                                          \\
		b2         & has 10 duplicates          & 1/0: are there at least 10 other nodes with the same text?                                                                                                              \\
		b3         & r same class path          & ratio of nodes on the page with the same class path (e.g. \texttt{body>div>a.link>b})                                                                                      \\
		b4         & has word                   & 1/0: there is at least one word in the text block                                                                                                                 \\
		b5         & log(n words)               & log(number of words) (clipped between 0 and 3.5)                                                                                                                        \\
		b6         & avg word length            & average word length (clipped between 3 and 15)                                                                                                                          \\
		b7         & has stopword               & 1/0: block contains a stopword                                                                                                                                          \\
		b8         & stopword ratio             & ratio of words the are in our stopword list                                                                                                                             \\
		b9         & log(n characters)          & log(number of characters) (clipped between 2.5 and 5.5)                                                                                                                 \\
		b10        & log(punctuation ratio)     & log(ratio of of characters $\in \{,, ?, ;, :, !\}$ to the total) (clipped between -4 and -2.5)                                                          \\
		b11        & has numeric                & 1/0: the node contains numeric characters                                                                                                                               \\
		b12        & numeric ratio              & ratio of numeric characters to the total character count                                                                                                                \\
		b13        & log(avg sentence length)   & log(average sentence length) (clipped between 2 and 5)                                                                                                                  \\
		b14        & ends with punctuation      & 1/0: the node ends with a character $\in \{,, ?, ;, :, !\}$                                                                                                             \\
		b15        & ends with question mark    & 1/0: the node ends with a question mark                                                                                                                                 \\
		b16        & contains copyright         & 1/0: the node contains a copyright symbol                                                                                                                               \\
		b17        & contains email             & 1/0: the node contains an email address                                                                                                                                 \\
		b18        & contains url               & 1/0: the node contains a URL                                                                                                                                            \\
		b19        & contains year              & 1/0: the node contains a word consisting of 4 digits                                                                                                                    \\
		b20        & ratio words with capital   & ratio of words starting with a capital letter                                                                                                                           \\
		b21        & ratio words with capital$^2$ & b25 squared                                                                                                                                                              \\
		b22        & ratio words with capital$^3$ & b25 to the power 3                                                                                                                                                       \\
		b23        & contains punctuation       & node contains a character $\in \{,, ?, ;, :, !\}$                                                                                                                       \\
		b24        & n punctuation              & number of characters $\in \{,, ?, ;, :, !\}$                                                                                                                            \\
		b25        & has multiple sentences     & 1/0: there are more than 1 sentences in the text                                                                                                                        \\
		b26        & relative position          & relative position of the start of this block in the source code                                                                                                         \\
		b27        & relative position$^2$        & 17 squared                                                                                                                                                              \\
		\hline
		b28        & has parent                      & 1/0: the CDOM leaf has a parent node                                                                                                                                    \\
		b29        & p body percentage          & ratio of the source code characters that is within the parent CDOM node                                                                                                 \\
		b30        & p link density             & ratio of characters within <a> elements to total character count                                                                                                        \\
		b31--b47    & parent features   & b6--b22, but for the parent CDOM node                                                                                                                                   \\
		b48        & p contains form element    & 1/0: the parent CDOM node contains a form element                                                                                                                       \\
		b49--b69   & aprent tag features        & encoding of the parent CDOM node's HTML tags as 1's and 0's \\
				   &                              &{\scriptsize(td, div, p, tr, table, body, ul, span, li, blockquote, b, small, a, ol, ul, i, form, dl, strong, pre)} \\
		\hline
		b69        & has grandparent                     & 1/0: the node has a grandparent CDOM node                                                                                                                                              \\
		b70--b89   & grandparent features       & b29--b48, but for the grandparent CDOM node                                                                                                                             \\
		\hline
		b90--b109  & root features              & b29--b48, but for the root CDOM node (body)                                                                                                                             \\
		\hline
		b110--b128 & tag features               & encoding of the CDOM node's HTML tags as 1's and 0's \\
		           &                             & {\scriptsize(a, p, td, b, li, span, I, tr, div, strong, em, h3, h2, table, h4, small, sup, h1, blockquote)}               \\
		\hline
		\end{tabular}}
		\vspace{0.1cm}\\
		\label{tab:block-features}
	\end{table}
	\FloatBarrier

\end{document}